\documentclass[10pt, twocolumn, unsortedaddress, superscriptaddress, raggedfooter, floatfix, showpacs, prl]{revtex4}
\usepackage{amsmath}
\usepackage{graphicx}
\begin{document}
\sloppy
\title{Muon Spin Relaxation Measurements on Zirconia Samples}
\author{C. Degueldre}
\affiliation{Laboratory for Materials Behavior, Paul Scherrer Institute, CH-5232 Villigen PSI, Switzerland.}
\author{A. Amato} 
\affiliation{Laboratory for Muon-Spin Spectroscopy, Paul Scherrer Institute, CH-5232 Villigen PSI, Switzerland.}
\author{G. Bart}
\affiliation{Laboratory for Materials Behavior, Paul Scherrer Institute, CH-5232 Villigen PSI, Switzerland.}
\begin{abstract}
Although of primary importance in the mechanistic understanding of Zircaloy hydriding, practically nothing is known on the transport properties of hydrogen in zirconia. In this frame the muon, which can be considered as a light hydrogen nuclide, can be used as an analogue and its behavior in zirconia and Zircaloy corrosion layer may provide more insight to understand the behavior of hydrogen in these phases. Preliminary muon spin relaxation ($\mu$SR) measurements on several monoclinic zirconia samples, including a Zircaloy corrosion layer, have been performed. From the observed muon depolarization rate, the muon diffusivity in bulk monoclinic zirconia can be extracted and is found comparable to that of recently reported proton diffusivity.
\end{abstract}
\maketitle

\section{Introduction}
The study on the behavior of hydrogen in zirconia for understanding its migration mechanism in the Zircaloy corrosion layers is justified as stated in specific studies and as reported recently (see Ref.~
\cite{ref1}):
\begin{quote}
Virtually, nothing is known about the form of hydrogen in ZrO$_2$ or how it migrates through it. (...) The arguments have always been over pre\-ci\-sely what changes are occurring in the oxide during this incubation time, and how the hydrogen migrates through the oxide.
\end{quote}

An atomistic picture of interstitial hydrogen in zirconia involves, as it does for hydrogen in dielectrics, determination of lattice sites, configurations adopted and the mechanisms of diffusion. An additional dimension is added in the study of muons in materials, namely the possibility of examining different states of the defects centers -- positive, neutral, negative -- respectively for the diamagnetic proton, the paramagnetic hydrogen atom and diamagnetic hydride ion. Muon spectroscopy has been successful in characterizing all these states, determining their local structures, their different stabilities, mobilities and interactions with charge carriers \cite{karlsson}. In addition characterization of trapping centers such as vacancies may be performed. Such similar studies could therefore be employed to get more insight on the behavior of hydrogen in for zirconia, as the muonium ($Mu = \mu^++e^-$) can be regarded as an isotopic analogue of the hydrogen atom, with a positive muon replacing the proton. 

In metals, the implanted muon comes at rest at an intersticial position and is in a diamagnetic state. After implantation, it may diffuse between interstitial sites and/or towards trapping spaces (defects, vacancies, foreign atoms). By $\mu$SR technique, information about the local magnetic field distribution at the muon site can be obtained. Such distribution will be sensitive to the dynamics of the local moments and/or of the muon itself. Consequently muon diffusion in metals may be investigated and numerous studies are reported for elemental metals (see for example Ref.~
\cite{borghini}). Alternatively, in dielectrics, part of the implanted muons can capture an electron and form a muonium pseudo-atom. The Muonium properties, including the nature of trapping sites and its diffusion, can in turn be studied by $\mu$SR, as succesfully demonstrated in a variety of studies (for a comprehensive review on muon and muonium diffusion see Ref.~
\cite{storchak}). 

As a complement of recent studies using x-ray absorption \cite{villela,conradson} to gain more insight on the influence of oxygen deficiency on electronic and local structure of monoclinic zirconia, the present $\mu$SR investigation aims to obtain complementary information on the muon diffusion on monoclinic zirconia (powder and  polycrystalline sample). Moreover, preliminary data on a Zircaloy corrosion layer are also reported.   

\section{Experiment}

\subsection{Muon Spectroscopy}

The muons used during $\mu$SR studies are basically 100\% polarized and are implanted one at a time into the sample, where they thermalize within few picoseconds. During his lifetime (ca. 2.2~$\mu$s), the muon can diffuse in the zirconia material as free (diamagnetic) muon or possibly as muonium. When the muon undergoes its weak decay a positron, among others, is produced which is preferentially emitted along the direction of the muon spin at decay time. For a $\mu$SR experiment, the direction along which the decay positron is emitted and the elapsed time interval between the muon implantation and the decay are determined. After $\sim 10^7$ muons have been observed, the obtained time histogram of the collected intervals has the form: 
\begin{equation}
N(t)  = N_0 \exp (-t/\tau _\mu ) [1+AG(t)] + Bg~{\text ,}
\end{equation}							
where $Bg$ is a time-independent background, $N_0$ is a normalization constant, $\tau _\mu$ is the muon lifetime and the exponential accounts for the muon decay. $AG(t)$ is often called the $\mu$SR signal. $A$ is the asymmetry parameter, which will depend on the experimental setup. 
$G(t)$ reflects the normalized muon-spin ${\mathbf S}$ autocorrelation function $G(t)=\langle{\mathbf S}(t)\cdot{\mathbf S}(0)\rangle/S(0)^2$, which depends on the average value, distribution, and time evolution of the internal fields sensed by the muons, and therefore contains all the physics of the magnetic interactions of the muon inside the sample \cite{blundell}. 
The present $\mu$SR measurements were obtained with the so-called transverse-field technique where a weak magnetic field is applied transverse to the initial spin direction. The form and width of the field distribution sensed by the muon is therefore detected by monitoring the decay of the muon polarization (i.e. reflected by the envelope $g(t)$ of the $G(t)$ function).

\begin{figure}
\includegraphics[width=8cm]{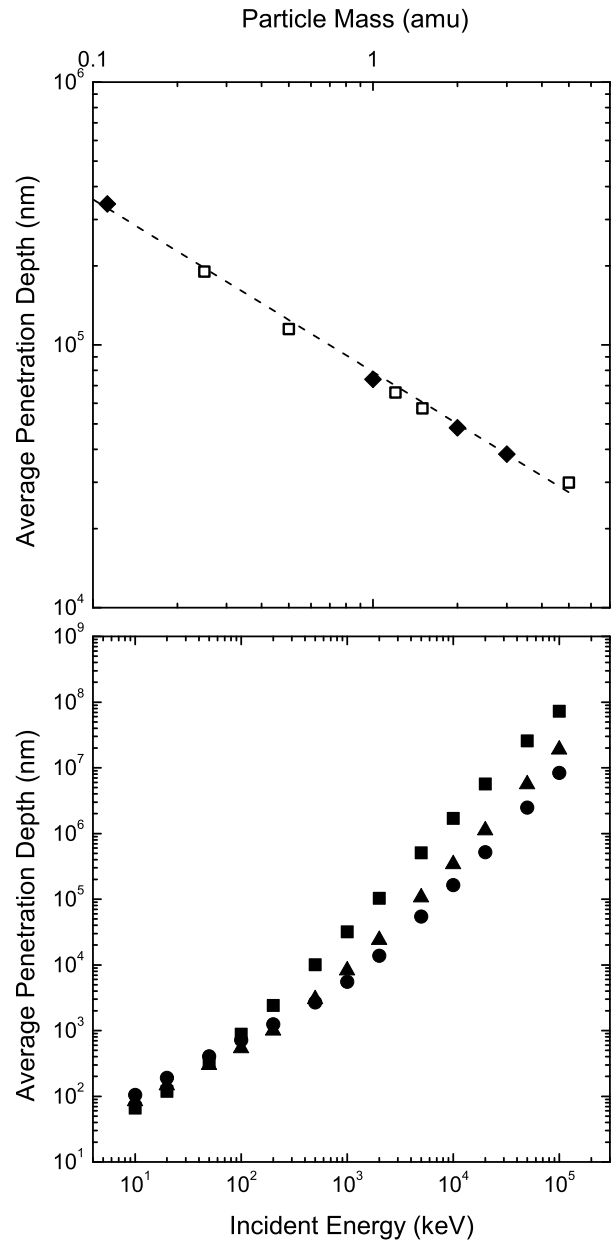}
\caption{\label{figure_1}Upper panel: Calculated penetration depth for leptons and hadrons with a 4~MeV incident energy (see text). Lower panel: Comparison between the calculated penetration depths of $\mu^+$ (squares), H$^+$ (triangles) and $^3$H$^+$ (circles).}
\end{figure}
The $\mu$SR experiments were performed at the Swiss Muon Source (S$\mu$S) of the Paul Scherrer Institute (PSI), Villigen (Switzerland). The GPS instrument \cite{gps} located on the surface muon $\pi$M3.2 beam-line was used. The typical range in matter of the muons available on this beamline is about 150~mg\,cm$^{-2}$. Samples were cryo-cooled in liquid helium and $\mu$SR was performed as a function of temperature between 5 and 300~K. Due to the small depolarization rates observed, the so-called Muon-On-Request (MORE) setup \cite{abela} was used. This setup dramatically reduces the value of $Bg$ and allows one to extend the usable $\mu$SR time window up to 20~$\mu$s and therefore detect depolarization rates as low as 0.001~$\mu$s.

\subsection{Samples}
Different monoclinic zirconia samples were investigated: i) a very pure powder sample; ii) a solid pellet sample; iii) a sample of natural monoclinic zirconia formed by a mosaic of single crystals. The  zirconia powder was produced by Fluka (pro-analysis) and was composed by particles in the size range of 1 to 10~$\mu$. The zirconia solid pellet was a sample of nanoparticles of pure zirconia pelletised and sintered at low temperature (kindly offered by the PennState University) as reported earlier by Raghavan \textit{et al.} \cite{raghavan}.  Natural zirconia was part of a brown rolled pebble comprising baddeleyite (ZrO$_2$) and zircon from Po\c{c}os de Caldas, Minas Gerais (Brazil). The baddeleyite part was $22.5 \times 22$~mm$^2$ with a thickness of 1.5 to 3~mm. This sample was kindly provided by the Mineralogical Department of the Geological Museum of the University of Lausanne (Switzerland).  The main impurities were Fe and Hf both at the percent level. In addition, a Zircaloy (Zry4) sample was also investigated. This sample ($40 \times 10$~mm$^2$), corroded in autoclave, was covered by a layer of ZrO$_2$. This monoclinic zirconia corrosion layer had a thickness of 100~$\mu$m as determined by gravimetry (1692 mg\,dm$^{-2}$). All samples were tested by x-ray diffraction and were found to assume the monoclinic structure of zirconia. 
\begin{figure}
\includegraphics[width=8cm]{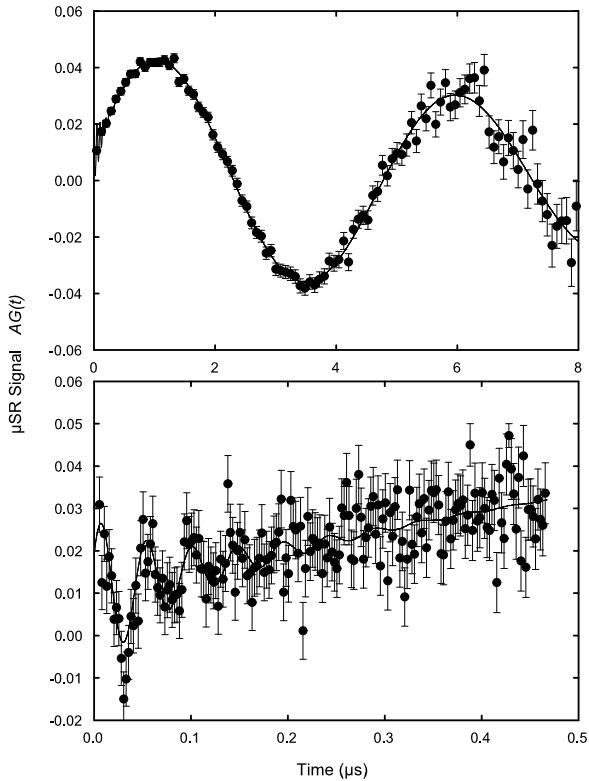}
\caption{\label{figure_zro2_raw} TF $\mu$SR spectra, recorded in an external field of 7~G over the full available time window (8~$\mu$s -- top) and over the first 0.5~$\mu$s (bottom), in monoclinic zirconia at low temperature (4~K). The signal surviving at long time is associated to diamagnetic muons, whereas the rapidly damped signal is ascribed to a muonium fraction (see text).}
\end{figure}

During $\mu$SR measurements, the samples were mounted either on special containers (powder sample) or on fork-like holders. In both cases, the fraction of muons missing the sample and creating a time-dependent background signal can safely be neglected.  

\section{Results and discussion}
\subsection{Modelling the Muon Implantation}

Penetration and implantation of leptons $^nx^+$ and hadrons $^nX^+$ in zirconia (density 5.9~g\,cm$^{-3}$) were modeled using the software \textit{SRIM2000} (Ref.~
\cite{srim}). First calculations were carried out for energy of 4~MeV corresponding to the muon energy available for the real measurements. The average penetration depth for the  $\mu^+$ amounts to 344~$\mu$m with a straggle of 23~$\mu$m.  For the same energy a proton penetration depth is 74~$\mu$m with a straggle of 2.3~$\mu$m. In this energy range, the deceleration mechanism occurs through electronic interactions. As expected, the heaviest particles have a shorter penetration depth. Figure~\ref{figure_1} illustrates this behavior for particles of various masses namely:  $\mu^+$, $^1$H$^+$, $^2$H$^+$, $^3$H$^+$, $\tau^+$, $\pi^+$ as possible particles for sample irradiation and $^{1.2}$H$^+$, $^{1.5}$H$^+$ and $^{5}$H$^+$ for hypothetical irradiation. It is striking to find a linear relationship on the log-log plot of the  penetration vs. mass. Additional modeling effort was focused to the comparison between penetration depths of  $\mu^+$, $^1$H$^+$ and $^3$H$^+$ at various energies. Figure~\ref{figure_1} (lower panel) shows a comparison between the three particles as a function of their energy. 

\subsection{$\mu$SR Results}
For all samples, transverse-field (TF) $\mu$SR studies were first performed in an applied magnetic field of 7~G. This low field value was choosen in order to differentiate the diamagnetic muon fraction from the muonium one (i.e. representing the formation of muonium in the sample). For all samples, a reduced fraction (typically 25\% of the total signal) of diamagnetic muons was observed at all investigated temperatures. The spin of these muons precesses according to the gyromagnetic ratio of the free muon [$\gamma_{\mu}/(2\pi)=13.55$~kHz\,G$^{-1}$ -- the muons in a muonium state exhibit a precession which is typically two orders of magnitude faster]. The remaining fraction of the $\mu$SR signal was related to the occurrence of muonium. However, a muonium signal could only be detected at very low temperature (below 6--7~K) where it is characterized by an extremely fast depolarization rate pointing to muonium delayed formation. Figure~\ref{figure_zro2_raw} exhibits a typical $\mu$SR spectra recorded at 4~K on monoclinic zirconia. Whereas diamagnetic muons are detected over the entire time window, neutral muonium is detected only at very early times (bottom panel of Fig.~\ref{figure_zro2_raw}). The population of prompt muonium is, however, rather low and reaches about 10\% at this temperature, A slight temperature increase has a devastating effect on the prompt muonium population and already at 6~K the population basically disappears ($< 1$\%), most probably in reason of delayed muonium formation It is consequently impossible already at low temperature to study the diffusion of muonium, even in a semi-quantitative way. Therefore, diffusion studies have been restricted to the diamagnetic muon fraction. For such TF $\mu$SR measurements with an external magnetic field of 50~G were performed. This method was preferred to the zero-field $\mu$SR technique since it allows the determination of the temperature evolution of very weak muon depolarization rates quite accurately.

\begin{figure}
\includegraphics[width=8cm]{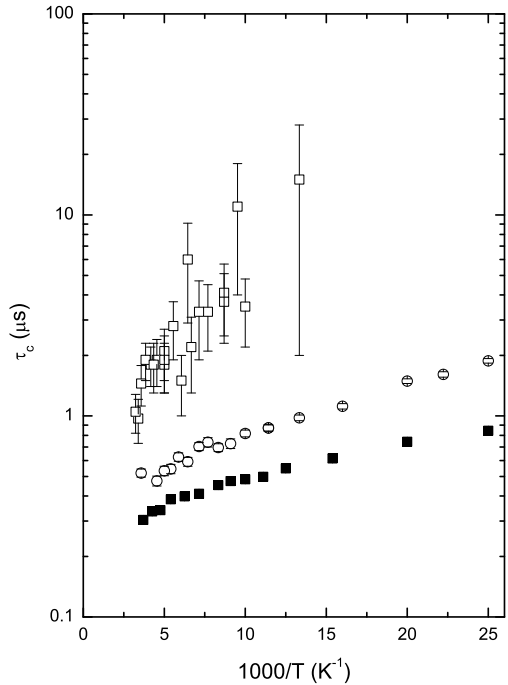}
\caption{\label{figure_zro2_arrhenius} Arrhenius plots of the parameter  $\tau_c$ measured different monoclinic samples of ZrO$_2$: in the powder sample (solid symbols), in the low-temperature sintered nanoparticle sample (pellet sample -- open squared) and in a natural sample (open disks). Note the increased errors bars for the nanoparticle sample due to the unavailability of the MORE setup for those measurements.}
\end{figure}
For all samples, the diamagnetic part of the $\mu$SR signal could satisfactory be fitted utilising the so-called Abragam formula, which describes the muon diffusion in transverse field (see for example Ref.~
\cite{schenck}):
\begin{equation}
\label{abragam}
G(t) = \exp [-M_2\tau_ c^2 \exp (-t/\tau _c)-1+t/\tau _c]\cos(\omega t + \phi)
\end{equation}
where $M_2$ is connected to the second moment of the field distribution at the muon stopping site ($M_2=\gamma_{\mu}^2\langle B^2\rangle$) inside the lattice and $\tau_c$ is the hopping time (which is related to the hopping rate $\Gamma=1/\tau_c$). The oscillatory term, on the right-hand side of Eq.~(\ref{abragam}) reflects the muon-spin precession with $\omega = \gamma_{\mu}B_i$ (where $B_i$ is the internal field at the muon site and of the order of the external field of 50~G) and $\phi$ is a phase essentially given by the geometry of the instrument. The Abragam function describes the time-dependence, introduced by the muon diffusion, of the interaction between the nuclear spins and the muon spin. If the muon motion is fast enough, the muon spin will sense an average nuclear field distribution, that is, the observed muon depolarization rate will decrease. In NMR the similar effect is known as motional narrowing. The Abragam function reduces, as expected, to the static depolarization function (Gaussian) in the limit of slow muon motion. On the other hand, for fast motion, the limit is correctly given by the exponential depolarization function and an overall decrease depolarization rate.

In the Eq.~(\ref{abragam}), the parameter $M_2$ was experimentally determined at low temperature, i.e. below about 20~K. In this temperature region, the depolarization of the diamagnetic $\mu$SR signal is found to be temperature independent and is interpreted as a regime with a slow hopping rate (i.e. $\Gamma=1/\tau_c \ll 1/\tau_\mu$ ), i.e. with a muon essentially static. On a second step $M_2$ was kept constant during all the remaining fitting procedure, corresponding to the assumption that the muon diffuses to equivalent sites. The parameter $M_2$ was found to equal 0.0256~MHz$^2$, 0.040~MHz$^2$ and 0.23~MHz$^2$ for the pellet, powder and natural sample, respectively. Such values are unexpectedly high in comparison to calculated field distribution at the possible muon site. Assuming the monoclinic structure and a perfect crystal, the field distribution at the muon site is produced by the nuclear magnetic moments, i.e. principally by the ion $^{91}$Zr$^{4+}$. This ion has a spin $I = 5/2$ with a moment $\mu=-1.298~\mu_{\text N}$ and a relatively low abundance of 11.2\%. For the most likely interstitial sites, and in particular for the site (0.5,0.5,0.5), theoretical calculations provide a parameter $M_{2\text{calc}}$ of the order 0.005~MHz$^2$. This huge discrepancy suggests that this specific material has inherently a large density of defects, impurities and/or vacancies. Consequently, the samples may contain important local distortions, not reflected in theoretical calculations. The major role played by defects on the magnetic field distribution at the muon site is illustrated by the rather high value of $M_2$ observed in the natural sample. In this sample, it is believed that a large contribution to the enhanced $M_2$ is provided by iron-rich clusters as FeO or Fe$_2$O$_3$.

Figure~\ref{figure_zro2_arrhenius} exhibits the parameter  $\tau_c$ obtained on the powder, pellet and natural samples by fitting Eq.~(\ref{abragam}) to the data. Clearly, the temperature evolution of the $\tau_c$ assumes an Arrhenius behaviour of the form:
\begin{equation}
\label{activation}
\frac{1}{\tau_c}=\nu_0\exp(-E_0/k_{\text B}T)~{\text ,}
\end{equation}
where $E_0$ is an activation energy and $k_{\text B}$ is the Boltzmann constant. Best fits of Eq.~(\ref{activation}) on the data provides activation energies of $3.6\pm0.3$~meV, $4.7\pm0.2$~meV and $16.2\pm0.9$~meV respectively for the powder, the natural and the pellet sample (see also table~\ref{table}). Eventhough the errors bars are rather higher, note that the hopping time $\tau_c$ extracted from the data obtained in the nanoparticle sample is rather longer than that of the other samples, corresponding therefore to a reduced hopping rate $\Gamma$. Such observation may be related to the presence of lattice strains leading to a reduction of the hopping probability through tunnelling.

From the value of $\tau_c$, and taking into account the nearly cubic symmetry of the zirconia system, the diffusion parameter $D$ can in principle be extracted by the expression\cite{schenck}:
\begin{equation}
\label{diffusion1}
D = \frac{1}{\tau_c}\frac{d^2}{6}~{\text .}
\end{equation}
Equation~(\ref{diffusion1}) assumes a perfect crystal, where the muon diffusion occurs between adjacent and identical interstitial stopping sites. Considering the most probable muon stopping site (0.5,0.5,0.5) mentioned above, the jump distance $d$ can be shown to be $d=a/\sqrt{2}$, where $a$ represents the lattice parameter. However, and as observed above, the picture of a clean and perfect crystal has to be questioned in view of the quite strong depolarization rate of the muon at low temperature (i.e. when the muon is essentially static). An alternative view is to consider that after thermalisation the muon comes at rest at a trapping site, which provides a deeper potential than a regular interstitial site. Such a scenario appears conceivable if the concentration of trapping sites is very high and/or if the `free' diffusion is extremely fast, meaning that all muons would be likely to have reached a trap during the first few nanoseconds. Equation~(\ref{abragam}) would still be valid, but now $\tau_c$ would represent the steps in diffusion between the trapping sites and $M_2$ would be connected to the second moment of the field distribution at a trapping site. As observed, such field distribution can be safely expected to strongly deviates (i.e. be wider) than the calculated one for normal interstitial sites in a perfect crystal. In this `dirty limit', and by assuming that the concentration of trapping sites is less than one site per unit cell, the determination of the diffusion parameter [see Eq.~(\ref{diffusion1})] can be considered as a lower limit. 
\begin{table}
\caption{\label{table} Comparison of the diffusion coefficient (estimated at 500~K) for various species in zirconia. The activation energy and species radius are also reported.}
\begin{ruledtabular}
\begin{tabular}
{lcccc}
            &   radius~(pm)   &   $D$~(m$^2$s$^{-1}$) &   $E_0$ (eV)   &   Ref.\\
\hline
Zr$^{4+}$   &   80        &   $10^{-20}$           &   --               &   
\cite{degueldre}\\
O$^{2-}$    &   136       &   $10^{-12}$           &   $3.2$            &   
\cite{degueldre}\\
Cs$^{+}$    &   167       &   $10^{-20}$           &  0.5--1.0          &   
\cite{degueldre}\\
H           &   78        &  $\sim10^{-16}$        &  --                &   
\cite{aufore}\\
H$^+$       &   --        &  $1.5\times10^{-12}$   &  --                &   
\cite{lim}\\
$\mu^+$     &   --        &0.5--1.0$\times10^{-13}$&3--16$\times10^{-3}$&   this work\\
\end{tabular}
\end{ruledtabular}
\end{table}

Table~\ref{table} presents a comparison of the diffusion coefficient $D$ obtained for various species in zirconia. The  values are estimated around 500~K using, if necessary, an Arrhenius law. The diffusion coefficients of ions such as O$^{2-}$ and 
Cs$^+$ in zirconia are reported in the literature \cite{degueldre}. At room temperature typical diffusion coefficients are found in the range $10^{-12}$ to $10^{-22}$~m$^2$s$^{-1}$ for O$^{2-}$ and Cs$^+$, respectively. The diffusion coefficient of hydrogen in zirconia (10$^{-16}$~m$^2$s$^{-1}$) was measured by Aufore \cite{aufore}. This study suggests a significant mobility in zirconia. On the other hand, Lim \textit {et al.} \cite{lim} studied by electrochemistry the mobility of proton in zirconia scales and found larger diffusion coefficient than for hydrogen. As already discussed, from the present study no information can be extracted for the muonium diffusion. However, an evaluation of the muon diffusion coefficient can be performed with the restrictions reported above concerning trapping effects. For sake of comparison, the coefficient $\tau_c$ was extrapolated at 500~K using Eq.~(\ref{activation}) and subsequently the lower bound of the diffusion coefficient $D$ was estimated using Eq.~(\ref{diffusion1}).

\begin{figure}[t]
\includegraphics[width=8cm]{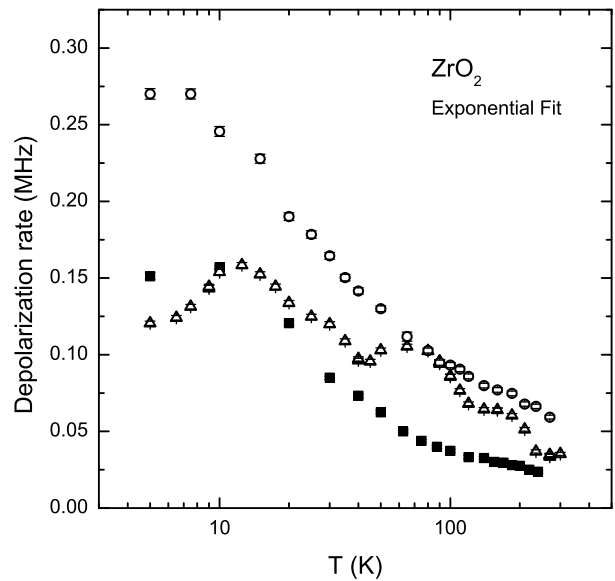}
\caption{\label{figure_layer}Comparison between the temperature evolution of the depolarization rate measured in different ZrO$_2$ samples. For sake of comparison, the data were here analysed with an exponential decay of the polarization (see also text). A natural sample (open circles), an ultra-pure powder (solid symbols) and a layer sample obtained from a corroded Zircaloy plate (open triangles). (Note that the error bars are smaller than the data points and that the layer sample data were performed during 4 different cooling and warming procedures with perfect reproducibility).}
\end{figure}
Finally, preliminary $\mu$SR studies have been performed on a monoclinic sample present as zirconia layer from a corroded Zircaloy sample. In view of the reduced thickness of the layer (100~$\mu$m) and of the available muon energy (about 4~MeV), a series of test runs have been performed to define the optimum degrader (460~$\mu$m). For the degrader, ultra-pure Al was utilized. The obtained thickness corresponds closely to the one calculated from the muon range (see above). As the temperature dependence of the $\mu$SR signal is strongly non-monotonic compared to the former samples, an analysis based on the Abragam formula [Eq.~(\ref{abragam})] is not possible and a simple exponential depolarization (reflecting satisfactorily the overall temperature dependence) was assumed. Figure~\ref{figure_layer} shows the $T$-dependence of the exponential depolarization rate of the corrosion layer sample compared to the former samples. The different maxima observed point to a number of trapping-detrapping effects occurring in the layer. Clearly, this suggests a variety of defects in the corrosion layer material. This variety impeeds a more quantitative analysis of the data obtained. Hence, each trapping site type $i$ can be characterized by a `mean time of stay' $\tau_0^i$ and the average time needed for a muon escaping from this kind of trap to reach the next trap variety $\tau_1^i$. The depopulation and repopulation processes lead to the characteristic humps observed in the temperature evolution of the muon depolarization rate. To complicate even more the picture, each type of trap has a given concentration $c_i$  and a given local field distribution $M_2^i$. By adding also the fact that the diffusion out of traps processes assume their own Arrhenius law (i.e. with different activation energy for the $\tau_0^i$), it is obvious that  quantitative information is, at least, extremely complex to extract. The present measurements should be basically taken as a strong hint for different trapping site types in the layer sample.

\section{Conclusion}
The first goal of this muon spectroscopy study was to provide more insight on the hydrogen and proton diffusion in zirconium oxide. The work was achieved by analysing muon behavior in zirconia samples. The muon interactions in monoclinic polycrystalline zirconia samples were investigated, with the aim of applying later the study on thin corrosion layers onto Zircaloy. Prompt muonium was not identified in zirconia above 6~K. The diffusion coefficient of muon in zirconia was estimated to be about $10^{-13}$~m$^2$s$^{-1}$ at around 500~K. In a prospective way the work has been completed by studying the properties of muon in a Zircaloy corrosion layer. 

\begin{acknowledgments}
This work was performed at the Swiss Muon Source (S$\mu$S) of the Paul Scherrer Institute (PSI), Villigen, Switzerland. Acknowledgements are due to Swiss Nuclear for its financial support of the LWV, to N. Meisser Curator of the mineralogical department Geological Museum of Lausanne for providing the natural zirconia sample and to S. Raghavan from University Penn state for providing the piece of pure monocline zirconia ceramic. The corroded Zircaloy sample was kindly provided by P. Barberis from CEZUS, Framatome-ANP, Ugine, France. The authors thank W. Hoffelner and U. Zimmerman for their interest in their work.
\end{acknowledgments}

\end{document}